# Pixelation-free and real-time endoscopic imaging through a fiber bundle

Donggyu Kim, Jungho Moon, Moonseok Kim, Taeseok Daniel Yang, Jaisoon Kim, Euiheon Chung and Wonshik Choi[*]

*Department of Physics, Korea University, Seoul 136-701, Korea*

**Endoscopy has been an indispensible tool in medical diagnostics,** and yet the demands for reduced unit diameter and enhanced spatial resolution have steadily been growing for the accurate investigation of distal sites with minimal side effects. However, the attempts to make use of thin image-guiding media accompany the degradation in spatial resolution as the micro-optics often induces aberrations. Here, we present a microendoscope that performs real-time correction of severe aberrations induced by image-guiding media such as a bundled fiber. Specifically, we developed a method exploiting the full binary control of a digital micro-mirror device (DMD) for characterizing the input-output response of image-guiding media and subsequently compensating the aberrations. As a proof-of-concept study, we completely eliminated the pixelation artifact, a severe form of aberration, in endoscopic imaging through an image fiber bundle and achieved spatial resolution much better than the diameter of an individual fiber. Our study lays a foundation of applying extremely thin, but highly aberrant image-guiding media for high-resolution microendoscopy.

Optical bio-imaging suffers from shallow imaging depth due to multiple light scattering occurring in biological tissues, which makes it extremely difficult to observe internal structures of specimens through skin tissues. Endoscopes have been developed in order to bypass this drawback, where an imaging unit with small diameter is inserted into human subjects and enables us to investigate internal sites in a minimally invasive manner. In recent years, the technology has been driven in a direction to reduce the diameter of imaging unit for relieving the degree of discomfort. For example, endoscopic imaging was performed by using an extremely thin single-mode optical fiber with a mechanical scanner and microlenses attached to the tip of the fiber[1]. In recent years, a bundled fiber has been widely used for the direct image delivery through single fibers in the bundle, sometimes in conjunction with confocal scanning at the input side of the bundle[2-6]. A graded-index (GRIN) rod lens and a needle-type microlens have also been used to perform confocal imaging[7,8] and optical coherence tomography (OCT) imaging at distal sites[9,10]. However, most of these approaches suffer from the degradation in spatial resolution because it is difficult to make micro-optics such as GRIN rod lens and microlens aplanatic as they become thinner. In the case of a bundled fiber, the pixelation effect, which is a form of aberration, is pronounced as the overall diameter of the bundle is decreased due to the reduction in the number of constituting fibers.

In recent years, various wavefront shaping methods have been developed for correcting severe aberrations and applied to endoscopic imaging through image-distorting media such as multimode optical fibers and highly scattering media[11-15]. The common underlying principle of the recently developed methods is to characterize the image distortion by measuring the medium's input-output response in the form of a transmission matrix or its equivalence. In our earlier study, we presented an endoscopic reflectance imaging method in which we employed an interference microscope equipped with a galvanometer scanning mirrors in the sample beam path to directly measure the transmission matrix[14]. We then eliminated image distortion by a post-processing based on the measured matrix. Recently, Mahalati et al. have also demonstrated reflectance imaging by using random speckle illumination and linear optimization algorithm[16]. As for the fluorescence imaging, Bianchi et al. and Cizmar et al. measured a transmission matrix by using interference of waves originating from different pixels in SLM[11,12]. They identified incident waves that would generate single focused spots at the opposite side of the fiber, which is a calibration process of the image distortion by the fiber. By shaping the incident waves with those calibrated ones, a focused spot was scanned at the sample plane to excite fluorophores. Similarly, Papadopoulos et al. used a digital phase conjugation to calibrate the multimode fiber[15,17]. Although these previous studies successfully demonstrated high-resolution microendoscopy through image-distorting media for reflectance imaging[14,16] and fluorescence imaging[11,12,15], their practical use for real-time endoscopic imaging was constrained due to either post-processing time or slow refresh rate of liquid crystal type of spatial light modulator (LC-SLM). In the work by Cizmar *et al.* in which LC-SLM and acousto-optic deflector were combined system to maximize the imaging acquisition rate, there was inevitable trade-off in the fidelity of the focus[11].

We note that the DMD can be a good alternative to the LC-SLM for the high-speed wavefront shaping purpose. It was previously used to generate a focused spot through a scattering medium by either a feedback control or the measurement of a transmission matrix for a single output point [18-20]. In the present study, we propose a novel method that makes full use of the binary control of DMD for recording the transmission matrix of a thin image-guiding medium such as a bundled fiber and a single multimode optical fiber. Specifically, we measured the complex field map at the output plane of the medium for individual micro-mirrors in the DMD. From the recorded transmission matrix, we identified a specific binary pattern that would generate a focus at an arbitrary spot at the far side of the medium. By changing the binary pattern at a refresh rate of 22,727 Hz, the full frame-rate of the DMD, we scanned the focus with the speed that is unmatched by LC-SLM, whose frame rate is at most 100 Hz. We achieved the image acquisition rate up to 10.3 Hz for $46 \times 46$ image pixels. In addition to the lateral scanning, we demonstrated axial focus scanning by applying numerical propagation algorithm to the measured transmission matrix. In doing so, we made the first experimental demonstration of real-time 3D fluorescence imaging through an extremely thin bundled fiber with the complete elimination of pixelation and spatial resolution close to the diffraction limit.

# RESULT

**Aberration calibration and pixelation-free focusing** Our experimental setup is schematically described in Fig 1. The backbone of the setup is an interferometric microscope with DMD installed in the sample beam path. From the interference between a reference plane wave and the wave transmitted through a fiber bundle, we recorded the complex field map of the field at sample plane (SP) located at 350 μm away from the bundle exit. The complex field at SP was the mixture of outputs from individual fibers, and became a fully developed speckle as a consequence. The DMD having numerous micromirrors (1024x768) was positioned at the conjugate plane to the input plane (IP) of the fiber bundle, and the binary control of DMD was used to eliminate pixelation artifact. In order to determine proper binary pattern of DMD, we measured the input-output response of the fiber, called a transmission matrix, $t$. The definition of $t$ depends on the choice of the basis. In our configuration, we considered spatial coordinates at IP and SP to be input and output bases, respectively. We set 7×7 micromirrors as a single macropixel, and $M$ = 10,000 macropixels constituted the input basis.

The element of the transmission matrix, $t_{ij}$, was determined as the complex field amplitude at the $i^{th}$ position in SP when a macropixel corresponding to the $j^{th}$ position in IP was turned on. A straightforward way to measure $t$ is to turn on each macropixel and then to record the complex field at the SP. However, this yields low signal to noise ratio (SNR) because only a small fraction of incident wave is used for the illumination. In order to maximize the SNR, we measured the superposed response of the fiber for the randomly chosen $M/2$ macropixels in the input basis, and then unfolded it to the response for each macropixel by means of the matrix inversion. To be specific, we first prepared for the $n$ = 12,000 binary sequences $S_{jp}$, ($j=1,2,\cdots,M; p=1,2,\cdots,n$ ) whose $M$ elements are either turned on or off with equal probability (Fig. 2(a)). These sequences represent the speckle basis. Then, we measured optical responses $O$, the complex field map at SP for the series of speckle basis at the IP (Fig. 2(b)). Using the measured *O* and predetermined *S*, we set the relation for the transmission matrix $t_{ij}$:

$$O_{ip} = \sum_j t_{ij} S_{jp}, \quad (i=1,2,\cdots,M; j=1,2,\cdots,N; p=1,2,\cdots,n). \tag{1}$$

Here *N* is the total number of output points at the SP. Then, the following relation calculated the transmission matrix:

$$t_{ij} = \sum_p O_{ip} S^{-1}_{pj}. \tag{2}$$

Here $S^{-1}$ stands for the inversion matrix of *S*. Since we measured the complex field map for each input basis by effectively $n/2$ times, we could increase SNR by $\sqrt{n/2}$ times higher than the simple method of turning on individual macropixels.

The recorded transmission matrix allows us to completely predict the transmitted wave at the SP for a given binary pattern of DMD. Moreover, we can apply for the numerical propagation of recorded *O* to calculate transmission matrix at a different depth in SP. In other words, the recorded transmission matrix at one specific SP is extended to the 3-dimensional space without additional measurement. As a result, we can not only compensate the effect of aberrations or distortion of imaging-guide media but also generate a diffraction-limited focus in 3D space of SP. In order to generate a focus at a target point *i'* at SP, we chose the complex conjugate of a row vector $t_{i'j}$ as a complex field map of input wave. In order to implement this input with the binary control of the DMD, we selectively turned on the input segments in which the phase of $t_{i'j}$ ranged between 0 and π (ref). Figure 2(c) shows the binary pattern displayed on DMD, which was designed to generate a focused spot at the center of SP. As shown in Fig. 2(d), a clean focused spot was generated at the target point as designed.

For the 10,000 segments as the number of input basis, the achieved enhancement factor of the focus, the ratio of the focus intensity to the background intensity, was 343. Though the enhancement factor is approximately half the value predicted by the theory[18], it is the best reported to date by the binary amplitude wavefront shaping method. The degradation of the focus fidelity comes from the cross talk between neighboring segments on DMD at the fiber IP, which reduces effective number of orthogonal segments. The spot size of the focus is determined by the speckle granular size at the SP, which is in turn related to the numerical aperture (NA) of the composite single fibers. The NA of the single fibers used in the experiment was 0.38. In order to assess the spot size, we have measured the point-spread function of a focused spot generated by the proposed method. We processed the measured transmission matrix and identified binary patterns that scan a focus along x-z plane at SP. While displaying the patterns on DMD, we acquired the intensity maps of the focus on x-z plane by using the camera (Fig. 3(a)). The line profile along lateral and axial directions are shown in Figs. 3(b) and 3(c), respectively. By fitting the data to the Gaussian function, we determined axial and lateral spot sizes, which were 8.99 μm and 1.07 μm, respectively. These are in excellent agreement with the theoretical values of 8.76 μm and 1.02 μm evaluated for 0.38 NA. This proves that the diffraction-limit focusing was achieved as well as the pixelation effect of the fiber bundle was completely removed.

**Pixelation-free and real-time fluorescence imaging** In order to perform fluorescence imaging, we scanned the focus at SP by displaying proper binary patterns found from the measured *t*. We could scan the focused spot at the frame rate of 22,727 Hz, which is the full frame rate of the DMD. The supplementary movie 1 shows a stream of intensity maps of focused spots taken while binary maps were displayed on DMD. With the scanning of the focus at SP ready, we performed fluorescence imaging for test objects and live biological cells. The experimental setup for the fluorescence endomicroscopy is embedded in Fig.1. The generated focus spot at SP excited fluorescence, which was then collected by the same fiber and propagated back to the input plane (IP). After reflected at the dichroic beam splitter (DS), the fluorescence was filtered by a long wavelength filter (LF) and a diffraction grating (DG), and then collected by the PMT. The detected signal was analyzed and the fluorescence images of the sample were reconstructed.

As a test sample, we used fluorescence beads (2 $\mu m$ skyblue, Spherotech). We dropped a solution containing the beads on a coverslip and waited for the solvent to dry such that beads were attached to the surface of the coverslip. The sample coverslip was positioned at SP. We first took a conventional endoscopic image of the sample through a bundled fiber for the purpose of comparison (Fig. 4(a)). In this case, we placed an LED (not shown) between DMD and a lens (TL1) to illuminate the sample from IP to SP. We then position an additional camera (not shown) at the focal plane of a lens (TL3) to record the reflection image of the sample. As can be seen in Fig. 4(a), the pixelation effect deteriorated fine details of the sample image. Then, we applied our method to acquire fluorescence endomicroscopy image (Fig. 4(b)). The effect of pixelation was completely removed and individual beads, whose size is smaller than the diameter of individual fibers, were clearly resolved. For this particular image, we scanned the focus at $70 \times 70$ target points covering the area of $58 \times 58$ $\mu m^2$ at SP, and image acquisition rate is 4.64 Hz. The supplementary movie 1 shows how the fluorescence image was unveiled as the scanning of the focus progressed. In terms of scanning speed, our method is more than 4 times faster than the recently reported method (ref). In order to demonstrate the real-time imaging capability, we took a stream of fluorescence images while moving the sample (Supplementary movie 2).

We also applied our method to taking fluorescence image of live biological cells. We stained human cancer cell-line (SNU-1074) by the fluorescence dye, SRfluor 680 Carboxylate (Polyscience) that stains peptides, proteins and antibodies. For the staining procedures, we prepared for the 1 mM solution of the fluorescence dye in DMSO. The solution was directly applied to the culture medium, and the cells were incubated for 5 minutes. Figure 3(c) shows the conventional endoscopic image of the cell taken by the fiber bundle. Figure 3(d) shows fluorescence image taken by our method, which is free from pixelation artifact. In addition, we demonstrated aberration-free endoscopic imaging through a single multimode optical fiber (Supplementary Fig. 1)

**Axial focus scanning and depth-selective fluorescence imaging** Our method is capable of performing 3D fluorescence imaging. Since the transmission matrix recording employed interferometric microscopy to take complex field maps at SP, we can numerically propagate each complex field to a the different depth at SP. Therefore, we can construct a new transmission matrix connecting the same IP, but to an output plane at different depths without measuring additional transmission matrix. In order to demonstrate 3D imaging capability, we performed 3D fluorescence images for fluorescence beads located at two different depths separated by 15 $\mu m$ (Fig. 5). From the transmission matrix measured at a reference depth, we calculated the transmission matrix at 15 μm farther from the reference depth. From those two matrices, we found the binary patterns that scans the focus at the two sample planes. Figures 5(b) and 5(d) show fluorescence images taken at the reference depth and propagated depth, respectively. One can notice that those beads blurred at one depth were clearly resolved at the other depth. For the comparison, we took bright field images of the sample in the transmission mode at the same foci as reference depth and propagated depth (Figs. 5(a) and 5(c), respectively). The camera recorded the intensity map of the sample image through an objective lens (OL2) while the LED illuminated the sample from IP side. We observed excellent agreement between the transmission images and endoscopic fluorescence images, confirming the 3D imaging capability of our technique.

**DISCUSSION**

Wavefront shaping technologies have been widely used to eliminate aberrations in optical imaging. They were successful enough to be practical in dealing with weak aberrations. In the case of severe aberrations, however, the techniques developed so far have been slow and impractical. The present study presented high-speed aberration compensation method for undoing severe image distortion. We believe that the proposed method has the potential to advance optical imaging modalities in the real practice. It will immediately resolve the drawbacks of using micro-optics in which aplanatic design is demanding. In the long run, the proposed method will lead to developing ultrathin endoscope with extremely high spatial resolution. While we have demonstrated microendoscopic imaging for the fluorescence, we can extend the method to other compelling imaging modalities such as two-photon imaging and OCT imaging.

**METHODS**

**Setup** The schematic layout of the experimental setup is shown in Fig. 1. The backbone of the setup is an interferometric microscope with DMD (D4100, Texas Instrument) installed in the sample beam path. The output beam from a He-Ne laser (25LHP928-230, CVI Laser Optics and Melles Griot, wavelength: 633 nm) was divided into sample and reference beams using a beam splitter (BS1). The sample beam was expanded to illuminate entire area of the DMD, and the wavefront of the reflected beam was shaped by selectively turning on and off its micro-mirrors. Then, the shaped pattern was delivered to the input plane (IP) of a fiber bundle (FIGH-06-300S, APL) by an objective lens (OL1, RMS40X, Olympus) and subsequently coupled to the fiber. The fiber bundle used in the experiment has 6000 ± 600 fibers and the outer diameter of the bundled fiber is 300 ± 25 $\mu m$. After propagating through the fiber, the transmitted speckle at the sample plane (SP) was delivered to the camera through an objective lens (OL2, RMS40X,

Olympus) and then interfered with the reference beam. The interference image was recorded by a complementary metal-oxide semiconductor (CMOS) camera (Motionscope M3 (500 fps), IDT), and then processed to acquire the amplitude and phase maps of the sample beam at SP. This complex field map of the output speckle was used to construct a transmission matrix of the fiber bundle. For the fluorescence imaging, we inserted a dichroic splitter (DS, T647lpxr, Chroma) at the upstream of the objective lens (OL1). A long pass filter (LP, FELH0650, Thorlabs) and a diffraction grating (830 Grooves, 800 nm NIR Ruled Gratings, Edmund) were used to filter the emitted fluorescence signal, which was detected by the photomultiplier tube (H5784-20, Hamamatsu).

**Recording transmission matrix** In the experiment, we assigned 7×7 micro-mirrors of DMD as an individual macropixel in input basis, and the total number of macropixels was 10,000. After 1/111 demagnification from the DMD to the IP, the size of a macropixel at IP was 860×860 nm$^2$. For speckle basis, we generated 12,000 random binary patterns, which 1.2 times larger than the number of macropixels, in order to increase SNR. The generated random patterns were uploaded to the onboard memory of the DMD and then displayed sequentially at a frame rate of 500 frames per second, limited by the frame acquisition rate of the CMOS camera. Figure 2(a) shows the representative binary patterns used in the experiment. For each random pattern, we recorded transmitted image at the camera (Fig. 2(b)). Using the set of images in Figs. 2(a) and 2(b), we obtained the transmission matrix of the fiber bundle based on Eq. (2). Using desktop computer operating with Intel i7-2600 CPU Quad Core (3.40 GHz), it took less than 5 minutes to obtain the transmission matrix.

**References**

1. Rivera, D. R. *et al.* Compact and flexible raster scanning multiphoton endoscope capable of imaging unstained tissue. *P Natl Acad Sci USA* **108**, 17598-17603, doi:Doi 10.1073/Pnas.1114746108 (2011).
2. Hopkins, H. H. & Kapany, N. S. A Flexible Fibrescope, Using Static Scanning. *Nature* **173**, 39-41, doi:Doi 10.1038/173039b0 (1954).
3. Kim, P. *et al.* In vivo wide-area cellular imaging by side-view endomicroscopy. *Nat Methods* **7**, 303-305, doi:Doi 10.1038/Nmeth.1440 (2010).
4. Koenig, F., Knittel, J. & Stepp, H. Diagnosing cancer in vivo. *Science* **292**, 1401-+, doi:Doi 10.1126/Science.292.5520.1401 (2001).
5. Flusberg, B. A. *et al.* Fiber-optic fluorescence imaging. *Nat Methods* **2**, 941-950, doi:Doi 10.1038/Nmeth820 (2005).
6. Matsunaga, T., Kawakami, Y., Namba, K. & Fujii, M. Intraductal biopsy for diagnosis and treatment of intraductal lesions of the breast. *Cancer* **101**, 2164-2169, doi:Doi 10.1002/Cncr.20657 (2004).
7. Lee, W. M. & Yun, S. H. Adaptive aberration correction of GRIN lenses for confocal endomicroscopy. *Opt Lett* **36**, 4608-4610 (2011).
8. Barretto, R. P. J. *et al.* Time-lapse imaging of disease progression in deep brain areas using fluorescence microendoscopy. *Nat Med* **17**, 223-U120, doi:Doi 10.1038/Nm.2292 (2011).
9. Tearney, G. J. *et al.* In vivo endoscopic optical biopsy with optical coherence tomography. *Science* **276**, 2037-2039, doi:Doi 10.1126/Science.276.5321.2037 (1997).
10. Li, X. D., Chudoba, C., Ko, T., Pitris, C. & Fujimoto, J. G. Imaging needle for optical coherence tomography. *Opt Lett* **25**, 1520-1522, doi:Doi 10.1364/Ol.25.001520 (2000).
11. Cizmar, T. & Dholakia, K. Exploiting multimode waveguides for pure fibre-based imaging. *Nat Commun* **3**, doi:Artn 1027
Doi 10.1038/Ncomms2024 (2012).
12. Bianchi, S. & Di Leonardo, R. A multi-mode fiber probe for holographic micromanipulation and microscopy. *Lab Chip* **12**, 635-639, doi:Doi 10.1039/C1lc20719a (2012).
13. Bertolotti, J. *et al.* Non-invasive imaging through opaque scattering layers. *Nature* **491**, 232-234, doi:Doi 10.1038/Nature11578 (2012).
14. Choi, Y. *et al.* Scanner-Free and Wide-Field Endoscopic Imaging by Using a Single Multimode Optical Fiber. *Phys Rev Lett* **109**, doi:Artn 203901
Doi 10.1103/Physrevlett.109.203901 (2012).
15. Papadopoulos, I. N., Farahi, S., Moser, C. & Psaltis, D. High-resolution, lensless endoscope based on digital scanning through a multimode optical fiber. *Biomed Opt Express* **4**, 260-270 (2013).
16. Mahalati, R. N., Gu, R. Y. & Kahn, J. M. Resolution limits for imaging through multi-mode fiber. *Opt Express* **21**, 1656-1668 (2013).
11

**Figure 1. Schematic experimental setup for recording a transmission matrix and fluorescence endomicroscopy**

An off-axis interference microscope with DMD installed in the sample beam path. The output beam from He-Ne laser was divided into sample and reference beams using a beam splitter BS1. Another beam splitter BS2 recombines the two beams. OL1, OL2: objective lenses, DS: dichroic beamsplitters, TL1, TL2, and TL3: tube lenses, LF: longwavelength filter, DG: diffraction grating, D: aperture, PMT: photomultiplier tube, IP: input plane of a fiber bundle, SP: sample plane 350 $\mu m$ away from fiber exit. The x, y, and z: spatial coordinates in SP. The magnification from DMD to IP was 1/111 and that from SP to camera is 55.6.

**Figure 2. Transmission matrix recording and diffraction-limit focusing**

(a) Representative maps of binary speckle basis, i.e. binary sequences $S$ in the main text, displayed on DMD. White (black) pixels indicate that mirrors are on (off) state. (b) Intensity maps at SP corresponding to the individual binary speckle basis in a. The $x$ and $y$ are the orthogonal coordinates at SP, and $p$ is the index of speckle basis. Scale bar, 10 $\mu m$. (c) A binary input pattern on DMD identified from the measured transmission matrix for focusing the laser beam at SP. (d) Output intensity image at the SP when the binary pattern in c was displayed on DMD. Scale bar, 5 $\mu m$.

**Figure 3. Point spread function of the focusing through a fiber bundle**

A focused spot was scanned along x-z plane and its intensity map was recorded by the camera. (a) Intensity map of the focus at x-z plane. The $z$ is the optical axis, and $x$ is one of the transverse coordinates in SP. Scale bar, 10 μm. (b) and (c) Line profiles along lateral and axial direction, respectively, in **a** and their respective full-width-half-maxima (FWHM) are 1.07 μm and 8.99 μm.

**Figure 4. Pixelation-free endoscopic imaging through fiber bundle**

(**a**) and (**c**) Conventional reflectance-mode endoscopic images of 2 μm fluorescence beads (Skyblue, Spherotech) and cancer cell line (SNU-1074), respectively, recorded through a fiber bundle. (**b**) and (**d**) Endoscopic fluorescence images of the same samples in **a** and **c**, respectively, recorded by the proposed method. The scale bar is 20 μm.

**Figure 5. Demonstration of the 3D endoscopic imaging**

(**a**) and (**b**) Conventional transmission-mode images of fluorescence beads taken at two different depths separated by 15 μm. Scale bar, 10 μm. (**c**) and (**d**) Endoscopic fluorescence images taken when the focus was scanned at the two corresponding depths of **a** and **b**, respectively.

Figure 1.

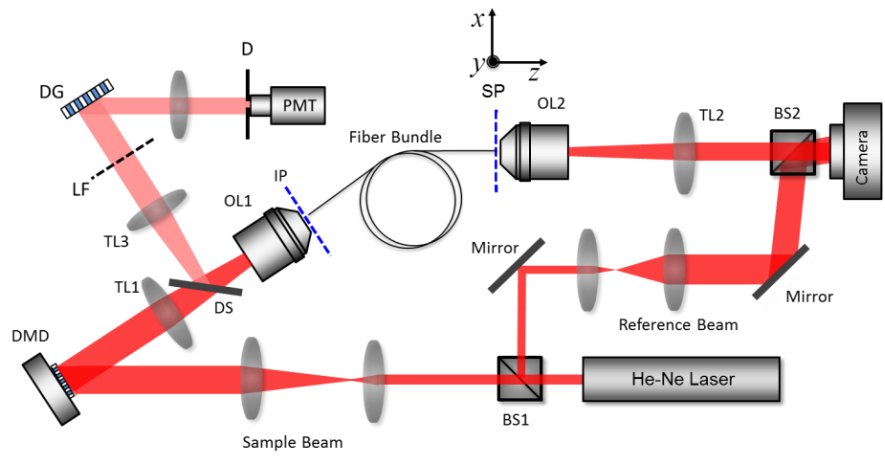

Figure 2.

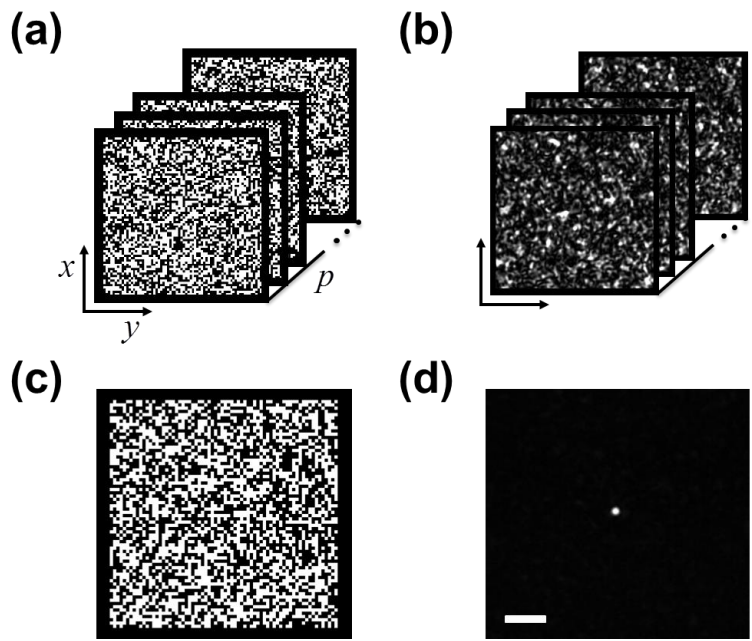

Figure 3.

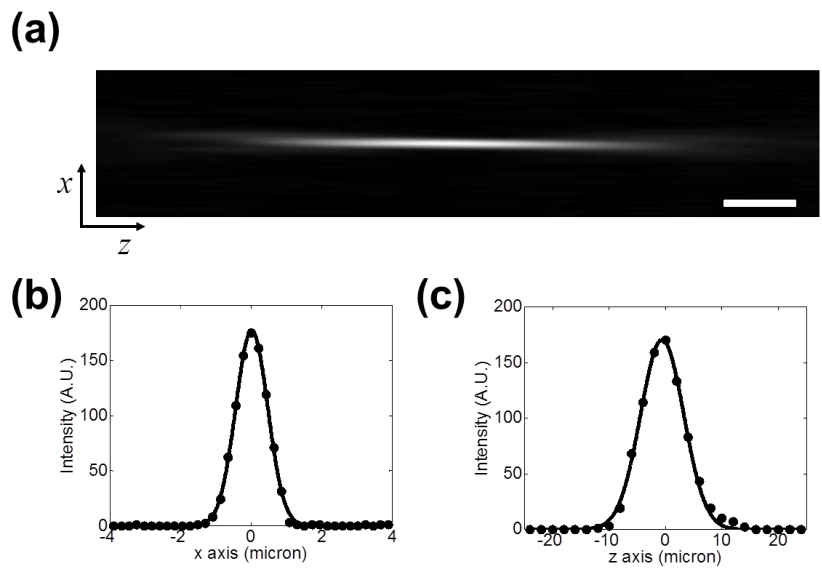

Figure 4.

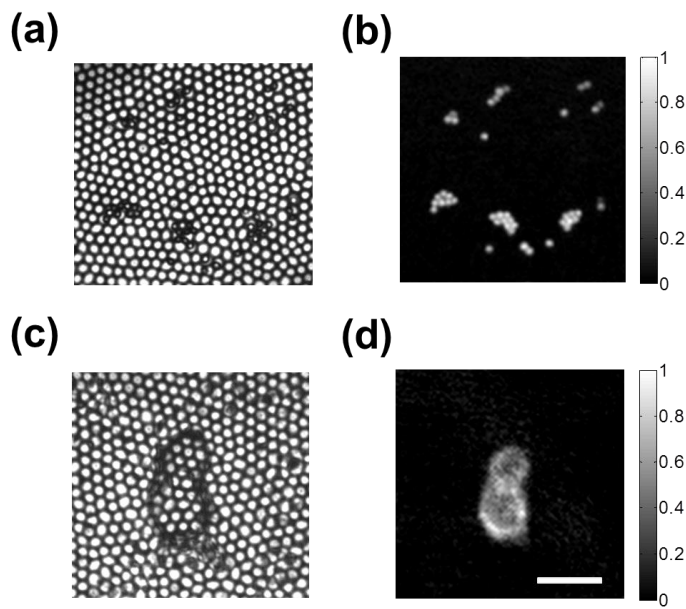

Figure 5.

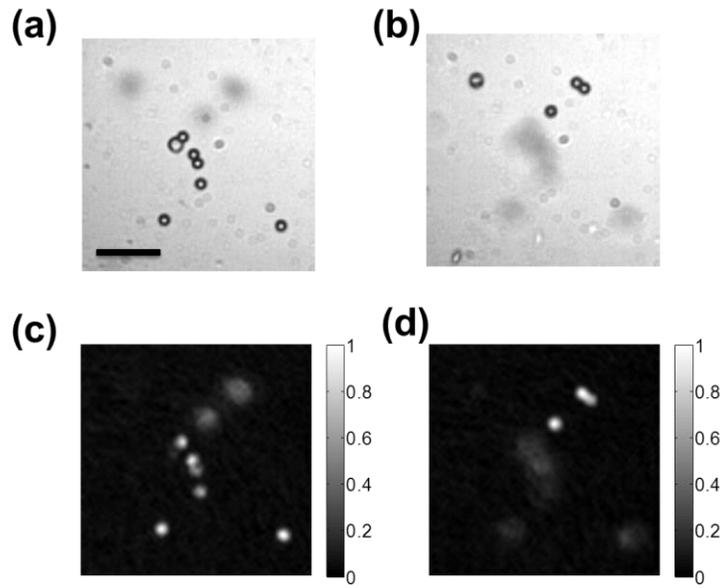